\documentclass[useAMS,usenatbib]{mnras}

\usepackage{amssymb}
\usepackage{amsmath}

\usepackage{epic}
\usepackage{epsfig}
\usepackage{rotating}
\usepackage{url}
\usepackage{xcolor} % for highlighting the changes in colour

\usepackage[T1]{fontenc}
\usepackage{ae,aecompl}

\title[Distant foreground and the Hubble constant]{Distant foreground and the \textbf{\emph{Planck}}-derived Hubble constant}
\author[V.N. Yershov, A.A. Raikov,  N.Yu. Lovyagin, N.P.M. Kuin, E.A. Popova]{ 
V.N. Yershov$^1$\thanks{E-mail: vyershov@list.ru}, A.A. Raikov$^{2,3}$,
 N.Yu. Lovyagin$^4$, N.P.M. Kuin$^5$, E.A. Popova$^3$  \\ 
{\small $^1$Corresponding address: MSSL, \mbox{}~Holmbury St.Mary, Dorking, Surrey, RH5 6NT, UK} \\   
{\small $^2$Special Astrophysical Observatory, Russian Academy of Sciences 
Niznii Arkhyz, 369167 Russia} \\
{\small $^3$Pulkovo Observatory, Russian Academy of Scences, 65-1 Pulkovskoye sh.,
Saint-Petersburg,  196140 Russia} \\ 
{\small $^4$Saint Petersburg State University, 7-9 Universitetskaya emb.,
Saint Petersburg, 199034, Russia} \\
{\small $^5$Mullard Space Science Laboratory, University College London, \mbox{}~Holmbury St.Mary, 
Dorking, Surrey, RH5 6NT, UK} 
}

\begin{document}

\date{Last updated 2020 February 17; in original form 2019 September 25} 

\maketitle

\begin{abstract}
It is possible to reduce the discrepancy between the local measurement of the cosmological 
parameter $H_0$ and the value derived from the {\it Planck} measurements of the Cosmic 
Microwave Background (CMB) by considering contamination of the CMB by emission from some 
medium around distant extragalactic sources, such as  extremely cold coarse-grain % 50
dust.  
Though being distant, such a medium would still be in the foreground with respect to the CMB,
and, as any other foreground, it would alter the CMB power spectrum. %80
This could contribute to the dispersion of CMB temperature fluctuations. %90
By generating a few random samples of CMB with different dispersions,  %101
we have checked that the increased dispersion leads to a smaller estimated value of $H_0$, 
the rest of the cosmological model parameters remaining fixed. This might explain the % 129
reduced value of the {\it Planck}-derived parameter $H_0$ with respect to the local 
measurements.  % 143
The signature of the distant foreground in the CMB traced by SNe was previously 
reported by the authors of this paper --
we found a correlation between the SN redshifts, $z_{\rm SN}$,  and CMB temperature fluctuations % 177
at the SNe locations, $T_{\rm SN}$. Here we have used the slopes of the regression lines %192
$T_{\rm SN}\,/\,z_{\rm SN}$ corresponding to different {\it Planck} wave bands in order % 201
to estimate the possible temperature of the distant extragalactic medium,
which turns out to be very low, about 5\,K. The most likely ingredient of this medium % 226
is coarse-grain  ({\it grey}) dust, which is known to be almost undetectable,
except for the effect of dimming remote extragalactic sources.  %247
\end{abstract}


\defcitealias{yershov14}{Yershov et al. (2014)}
\defcitealias{yershov12}{Yershov et al. (2012, 2014)}
 
\section{Introduction}

The first and consecutive releases of the {\it Planck} mission results revealed a 
statistically significant discrepancy between the  
cosmological parameter $H_0$ as calculated by using the {\it Planck} measurements
of the Cosmic Microwave Background (CMB) radiation
within the standard $\Lambda$ cold dark matter ($\Lambda$CDM) 
cosmological model, $H_0=(67.37\pm0.54)$%}%
\,km\,s$^{-1}$\,Mpc$^{-1}$ \citep{{aghanim18}},
and the values of this parameter obtained by using other methods -- 
mostly from direct local measurements \citep[see the review by][]{riess20}. 
One of these local measurements is based on optical and infrared (IR)
observations of variable Cepheid stars, 
with the recent calculation of $H_0$ based on this method being 
$H_0=(73.48\pm1.66)$\,km\,s$^{-1}$\,Mpc$^{-1}$ \citep{riess18}.

Both local and {\it Planck}-derived estimates of $H_0$ have passed a number 
of rigorous tests by considering many possible sources of systematic errors
\citep{efstathiou13, zhang17, planck17, feeney18, follin18}, but the discrepancy
still remains, which has resulted in active discussions 
in the literature.
Most authors seek to explain this discrepancy by implicating either unknown systematic effects 
in the observations or by focusing primarily on the possibility of new physics 
beyond the standard cosmological model and\,/\,or beyond the
standard model of particle physics. 

Possible modifications of the $\Lambda$CDM model include a new
kind of dark energy \citep{ade16d, ruiyun18} or  
an increase of the number of parameters 
in this model -- for example, from 6 to 12, as was proposed by 
\cite{valentino16}.
There are many other exotic proposals, like decaying dark matter, dark radiation,
modified gravity etc., 
most of which are based on some departure from basic cosmological principles.

On the particle physics side, 
there are advocates for the existence of new relativistic particle species, such as
hypothetical sterile neutrinos
\citep{wyman14,dvorkin14,sakstein19}, which could lead to a smaller 
expansion rate at early times and, thus, explain the $H_0$ 
discrepancy. But this is at the expense of departing from the standard model of particle physics.    

It is worth while mentioning that the baryon acoustic oscillation probe based on
galaxy surveys combined with big bang  nucleosynthesis data, gives an independent estimate of 
$H_0=68.3^{+1.1}_{-1.2}$ \,km\,s$^{-1}$\,Mpc$^{-1}$ \citep{schoneberg19} which is 
close to the {\it Planck}-derived $H_0$ value. These are important results, but discussing them 
goes beyond the scope of this paper.

\citet{valentino19} have found that the $\Lambda$CDM parameters derived from the {\it Planck} data 
can be brought into agreement with each other and with the Cepheid-based $H_0$ value  
by assuming a closed Universe with the curvature parametrised by the 
energy density parameter $\Omega_K=-0.091 \pm 0.037$. According to these authors,
such a fit requires drastic changes in the $\Lambda$CDM model, unless there exists       
an obvious possible solution to this problem in the form of 
hitherto undetected systematics in the CMB data.

In our view, it is preferable to find a solution 
to the $H_0$ discrepancy problem from within more conventional physics,
perhaps by using some additional observational data and\,/\,or re-examining the interpretation
of the existing measurements in terms of biases and systematics.    
The main systematic effects in the {\it Planck} data (e.g.  
the Galactic foreground radiation) are removed 
during the pipeline processing \citep{ade16a}, whereas the residual effects
are analysed and removed by simulations and numerical modelling of the instruments 
\citep{ade16b}. All the procedures for removing systematic effects have been rigorously
checked and validated. 
Extragalactic foregrounds are usually modelled by a set of power spectrum 
templates \citep{ade16c} that include fluctuations in the number density and 
clustering of extragalactic point sources, as well as the thermal and kinetic 
Sunyaev-Zeldovich components. 

However, despite taking into account all these sources of 
contamination, the lower value of $H_0$ derived from the {\it Planck}
measurements persists in contrast to the local measurements.
This means that there might still exist an
additional component (or various components) contaminating the CMB.

Here we shall explore such a possibility 
based on our previous work \citep{yershov12,yershov14} which revealed  
that the  CMB temperature fluctuations could be affected by emission from 
the medium around remote clumps of matter (e.g. galaxy clusters or superclusters). 
This conclusion came from finding a statistically significant correlation between the 
SN redshifts, $z_{\rm SN}$,  and CMB temperature fluctuations 
at the SNe locations, $T_{\rm SN}$. 
In this paper we compare and contrast our predicted change in the CMB power 
spectrum and in the estimated value of $H_0$ due to the increase in the 
dispersion $\sigma_T$ of contaminated CMB temperature fluctuations $\Delta T$
-- contaminated with respect to
the theoretically clear case of the standard $\Lambda$CDM model.

\section{Possible distortions of the CMB power spectrum}

In order to quantify the corresponding changes in the CMB power spectrum we have used 
the code for anisotropies in the microwave background (CAMB)  
 created by \citet{lewis13}  
in its 2014-version from {\it github}% 
\footnote{\url{https://github.com/cmbant/CAMB/releases/tag/Apr2014}}, 
which allows extracting different cosmological parameters 
from theoretical CMB power spectra generated by the same code.

\begin{figure}
%\begin{tabular}{l}
\hspace{-0.5cm}
\epsfig{figure=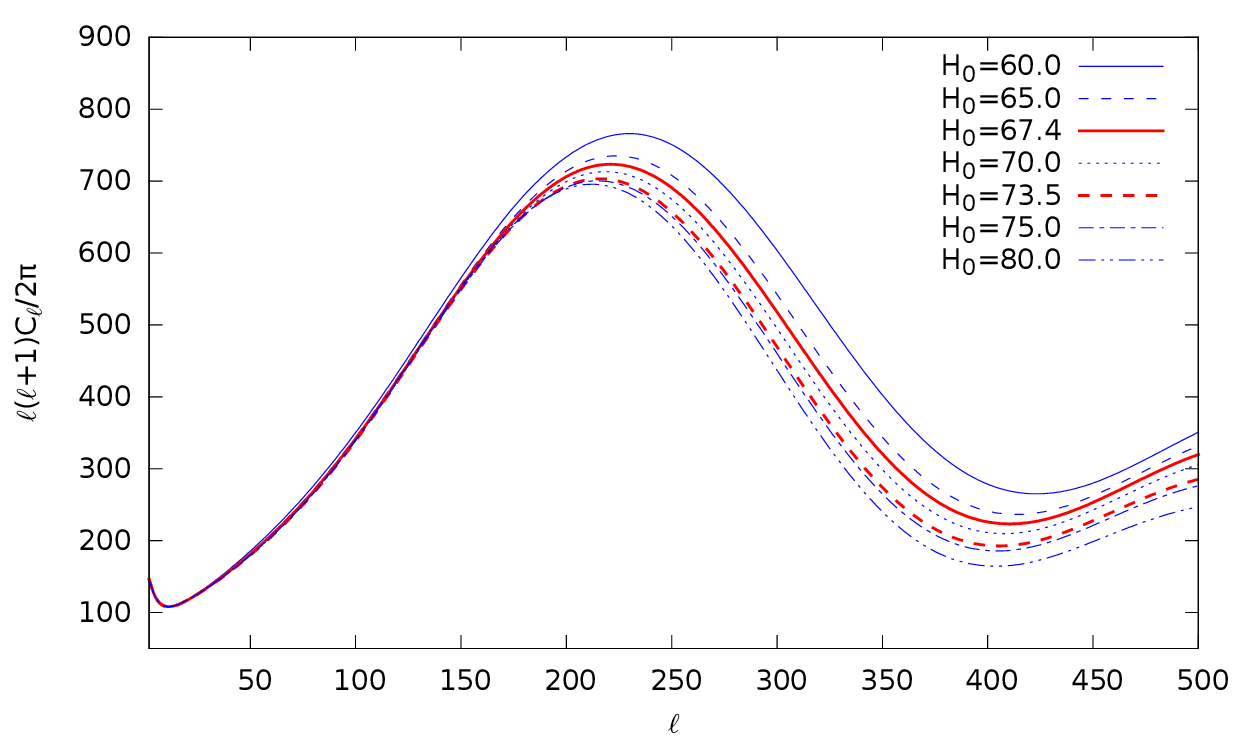,width=9.0cm} \\
\caption{CMB power spectra (in the standard normalised presentation) generated by using 
the CAMB tool for seven different values of $H_0$. 
} 
\label{fig:h}
\end{figure}

By fixing the parameters $\Omega_\Lambda$ and  $\Omega_M$
and varying the value of the parameter $H_0$ we have  
produced a set of theoretical CMB power spectra, part of which is 
shown in Figure~\ref{fig:h}.  
The fixed parameters have been given the standard default 
values of $T_0=2.7255$\,K, $\Omega_{\rm baryon}=0.0462$, $\Omega_{\rm CDM}=0.2538$ and $\Omega_\Lambda=0.7$
(which corresponds to $\Omega_{\rm M}= \Omega_{\rm baryon} + \Omega_{\rm CDM} = 0.3$ for the flat Universe).

In this code, the coefficients  $C_\ell$ of the CMB power spectrum are calculated as 
sums of the integrals $a_{\ell m}$, $|m| \le \ell$, which include
temperature fluctuations $\Delta T(x,\phi)$ over the celestial sphere, where
$x\in[-1,1]$ is the cosine of the latitude and $\phi\in[0,2\pi]$ is the longitude.
Reversely, the functions $\Delta T(x,\phi)$ are calculated by summing up the integrals
$a_{\ell m}$.

For a given CMB power spectrum, we have calculated a set of corresponding values of 
$\Delta T(x,\phi)$
over a 1/1024th part of the celestial sphere
by using random $a_{\ell m}$ for 
$\ell=0,1,\dots,\ell_{\tt max}$ with the 
restriction $\ell_{\tt max}=500$ 
(the use of the whole sphere is not needed here, as the angular sizes of distant cosmic 
structures in question are not expected to be large).
In these calculations,  we have taken five equal-spaced values of $H_0$, namely, 
60, 65, 70, 75 and 80 [km s$^{-1}$ Mpc$^{-1}$]  
plus the values 67.4 and 73.5 [km s$^{-1}$ Mpc$^{-1}$] 
corresponding, respectively, 
to the {\it Planck} result \citep{aghanim18} and to the local measurements of $H_0$ \citep{riess18}.
Figure~\ref{fig:h} shows the CMB power spectra generated in this way.

Additionally, for checking the consistency of our calculations we 
have taken a few sets of normally distributed randomised values of
$a_{\ell m}^i$, $i=1,2,\dots,5$, so that for each of the selected 
values of $H_0$, we have obtained five samples of $a_{\ell m}^{H_0,i}$ and, 
correspondingly,  five samples of values $\Delta T^{H_0,i}$.
For each of them, we have calculated the average of the CMB temperature 
fluctuations $\overline{\Delta T}$ and its standard deviation $\sigma_{T}$.
Here we are mainly interested in the way the values $\sigma_T$ change 
when the parameter $H_0$ is varied.  
For each of these generated sequences, the trend of the 
calculated values $\sigma_{T}$ was practically the same.
Namely, when the dispersion of the CMB temperature fluctuations increases, the 
value of the estimated $H_0$ diminishes, the difference between the two 
discussed $H_0$ values 73.5 and 67.4  [km s$^{-1}$ Mpc$^{-1}$] being related to
$\Delta\sigma_{T}=-0.60\pm 0.04\,\mu$K.
This supports
our proposition that if the CMB is contaminated 
by photons from the medium surrounding remote clumps of matter,
then the {\it Planck}-derived parameter $H_0$ extracted from such contaminated
CMB data would be underestimated with respect to the same parameter 
derived from a theoretically clean CMB case or from 
local observations not related to the CMB.

%%%%%%%%%

\section{Contaminating emission in different frequency bands}

Let us estimate the possible range of temperatures of the medium
from which the contaminating photons might emanate.   
In our previous work, \citetalias{yershov14}, we confidently excluded the 
possibility of the correlation between the SN redshifts and the 
CMB temperature fluctuations being caused by the Sunyaev-Zeldovich (SZ) effect
by comparing the signs of the correlation in different frequency bands. 
The SZ effect should cause a decrease in the CMB intensity at  
frequencies below 218 GHz and an increase at higher frequencies.
Therefore, if the observed anomaly was caused by this effect,
we would expect a higher positive anomaly  for the 353 GHz band.
However, the effect was exactly the opposite: the anomaly 
in the 353 GHz band was negative, and the slope of the 
regression line for this band was negative as well, 
$\xi_{353}=-61.8 \pm 30.0~[\mu{\rm K}]$.
The remaining possibility is the existence of contaminating emission  
from a cold medium residing in and around the SN host galaxies.

\begin{figure}
\hspace{-0.5cm}
\epsfig{figure=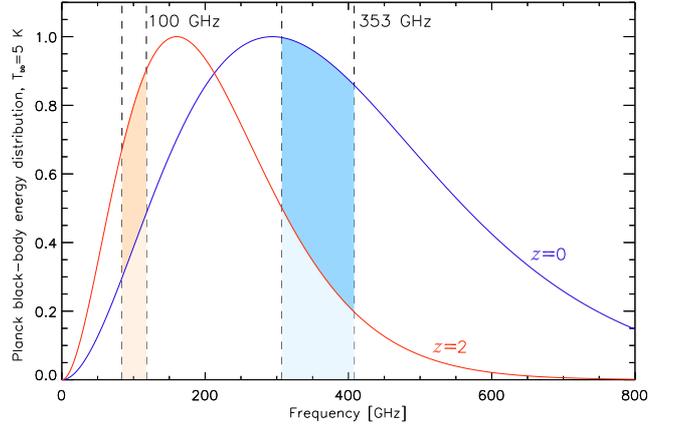,width=9.0cm}
\caption{Changes in the blackbody energy fraction observed in two {\it Planck}
 HFI bands (100\,GHz and 353\,GHz) due to redshifting the 
radiating matter with $T_{\rm bb}=$5\,K from  $z=0$ (blue curve)
to $z=2$ (red curve); the observed energy is redistributed from the band 353 GHz (blue shaded
area) to the 100 GHz band (salmon shaded area); the Planck blackbody curves are normalised 
here to their maxima;  the Planck blackbody 
curves are normalised here to their maxima; 
the thermal equilibrium threshold corresponding to the CMB temperature at $z=2$
is taken into account.} 
\label{fig:planck_radiance100_353}
\end{figure}

As an illustration, let us calculate the Planck blackbody energy distribution 
for the photons coming from some cold ($T_{\rm bb}=$5\,K) medium 
at two different redshifts, as shown in Figure~\ref{fig:planck_radiance100_353}.
The blue and red curves in this plot correspond to $z=0$  and $z=2$, respectively.  
The dashed vertical lines indicate the average limits
[84.4, 117.36]\,GHz and [306.8, 408.22]\,GHz
of the 100\,GHz and 353\,GHz {\it Planck} frequency bands, respectively
\citep{ade14spresp}. 
Of course, the real shapes of the {\it Planck} band transmission curves are not 
rectangular, 
so for our further estimations we have used the actual transmissions available 
at the ancillary-data folder of the {\it Planck} legacy archive   
\footnote{\url{https://irsa.ipac.caltech.edu/data/Planck/release_1/}}.

Figure~\ref{fig:planck_radiance100_353} schematically illustrates the effect of 
redistribution of photons from the band 353\,GHz into the band 100\,GHz 
if these photons are redshifted to $z=2$ (for simplicity the Planck blackbody curves in this plot are 
normalised to their maxima). The fraction of the blackbody photons observed in the instrumental
band of 353 GHz and coming from  $z=0$ (blue-shaded area in the plot) becomes essentially smaller
if the photons with the same blackbody temperature are redshifted
to $z=2$ (light-blue shaded area) because these photons are now observed in the  100 GHz band
(salmon-shaded area in the plot, which is smaller for this band at $z=0$ and which is shaded 
with light-salmon colour). This would lead to a positive slope of the 100\,GHz-band signal with 
increasing redshift and to a negative slope for the 353\,GHz-band signal.
In Fig.~\ref{fig:planck_radiance100_353}, we have taken into account the temperature threshold for a medium in thermal equilibrium 
with the CMB whose temperature at $z=2$ is higher than 5\,K, which shifts the peak of the red curve 
to higher frequencies, compared to the pure 5\,K-blackbody redshifted to $z=2$.

\begin{figure}
\hspace{-0.5cm}
\epsfig{figure=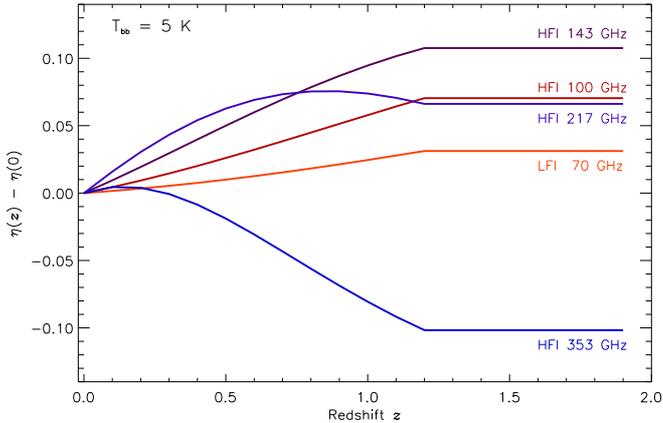,width=9.0cm}
\caption{Redshift dependences of the blackbody radiation energy
fraction $\eta(z)$ as observed in five {\it Planck} 
frequency bands for $T_{\rm bb}=5$\,K; the constant shifts of the curves
with respect to each other have been normalised at $z=0$ by 
subtracting from them their individual values $\eta(0)$.} 
\label{fig:bb_slopes}
\end{figure}

In this way, for this particular blackbody temperature of $T_{\rm bb}=5$\,K,
we can calculate the fraction of photons, $\eta(z)$, detected 
in each instrumental frequency band for a range of redshifts of interest.
The results of such a calculation are shown in Figure~\ref{fig:bb_slopes} for the 
five selected frequency bands of 70, 100, 143, 217 and 353 GHz and for the 
redshifts ranging from $z=0$ to $z=2$. Here, the threshold corresponding
to thermal equilibrium with the CMB at different redshifts is also taken 
into account, which can be seen in the shapes of the curves.
The linear regression coefficients (slopes) $\eta(z)$ corresponding to these curves  
indicate how the redshift dependence would look in different frequency 
bands if the contaminating emission were coming from different redshifts.
Since here we are interested only in the slopes
of these curves and not in their absolute values, 
we have normalised their constant shifts with respect to each other 
at $z=0$ by subtracting from them their
individual values $\eta(0)$.

\begin{figure}
\hspace{-0.5cm}
\epsfig{figure=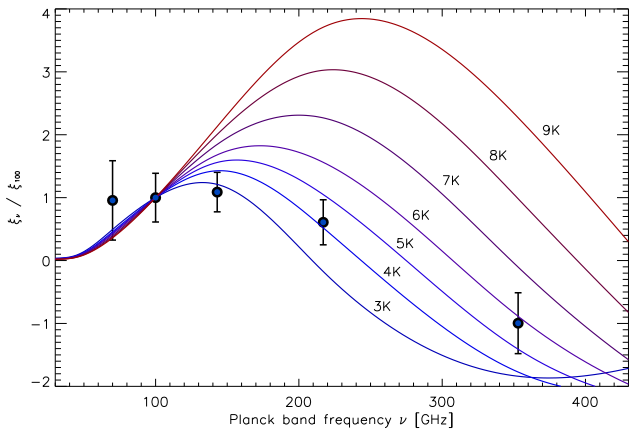,width=9.0cm}
\caption{Theoretical (solid curves) and observed (points with error-bars) slopes 
of the regression characterising the relationship between $T_{\rm SN}$ and $z_{\rm SN}$ for different  
{\it Planck} frequency bands; the slopes are normalised to the slope magnitude for the HFI 100 GHz band.} 
\label{fig:bb_observed_slopes}
\end{figure}

By comparing Figure~\ref{fig:bb_slopes} with our previously published results, namely, 
with figure~3  and table~5 from \citetalias{yershov14}, we can see that
the pattern of slopes of the calculated functions $\eta(z)$ for the 
five selected {\it Planck} frequency bands matches the pattern of the observed slopes 
of the functions $T_{\rm SN}(z)$ for the same frequency bands.   

The calculated theoretical slopes of $\eta(z)$  and the experimental slopes of  
$T_{\rm SN}(z)$ are expressed in different units:  the former is 
a dimensionless ratio between the blackbody energy integrated within the limits  
of a given {\it Planck} band and the total blackbody energy, whereas the latter is expressed in 
Kelvins per unit redshift.    
Therefore, in order to be able to compare these functions 
we have to normalise them by choosing a reference curve corresponding to one of the frequency bands, 
say, 100 GHz. The theoretical slopes $\xi_\nu$ of the functions 
$\eta(z)$ calculated in this way for different {\it Planck} bands, normalised to the slope of the 
function $\eta(z)$ for the band 100 GHz, are given in Table~\ref{theoretical_slopes} 
for temperatures of 3\,K, 4\,K, 5\,K, 6\,K and 7\,K. 
The measured experimental slopes taken from table~5 of \citetalias{yershov14}, also
normalised to the experimental slope corresponding to the 100 GHz band, 
are given in the last column of Table~\ref{theoretical_slopes}. 
The data from Table~\ref{theoretical_slopes} are illustrated  by 
Figure~\ref{fig:bb_observed_slopes}, where the experimental slopes are 
shown as black points with error bars.

\begin{table}
 \caption{Theoretical slopes $\xi_\nu$ of the function $\eta_\nu(z)$ 
normalised to the values of $\xi_{100\,{\rm GHz}}(z)$ for different 
temperatures of the blackbody emitter and the experimental 
slopes derived from the {\it Planck} data 
taken from \citetalias[table 5]{yershov14}.}
 \label{theoretical_slopes}
 \begin{tabular}{@{}lrrrrrr}
  \hline
  Band &   \multicolumn{5}{c}{Slopes for the blackbody temperatures}   & Experimental    \\
  GHz  & \hspace{-15pt} 3\,K & 4\,K  & 5\,K  & 6\,K  & 7\,K  & slopes   \\
  \hline
  70  & \hspace{-15pt} 0.49    & 0.46     & {\bf 0.44}     
  & 0.42   &  0.39 & $\mathbf{0.96}\pm 0.63$ \\
  100 & \hspace{-15pt} 1.00    & 1.00     & {\bf 1.00}     & 1.00   &  1.00 & $\mathbf{1.00}\pm 0.39$ \\
  143 & \hspace{-15pt} {\bf 1.21} &  1.42     & {\bf 1.56}  
     & 1.67   &  1.84 & $\mathbf{1.09}\pm 0.31$ \\
  217 & \hspace{-15pt} $-0.17$ & {\bf 0.52}   & {\bf 1.04}
       & 1.56   &  2.26 & $\mathbf{0.61} \pm 0.36$ \\ 
  353 & \hspace{-15pt} $-1.84$ &  $\mathbf{-1.77}$ & $\mathbf{-1.41}$ 
  & $\mathbf{-0.89}$ & $-0.10$ & $\mathbf{-0.99}\pm 0.48$ \\
  \hline
 \end{tabular}
\end{table}

We can see that these experimental slopes match various theoretical 
slopes corresponding to the blackbody emission with temperatures
from 3\,K to 6\,K, different temperature columns matching $1\sigma$-experimental
tolerances for different {\it Planck} bands, and all of the 5\,K-column slopes
matching all of the experimental slopes within $\sim 1.1\sigma$-tolerance
(hence the choice of 5\,K for the illustrations in
Figure~\ref{fig:planck_radiance100_353} and Figure~\ref{fig:bb_slopes}). 
This temperature is the lowest possible for a medium in 
thermal equilibrium with the CMB radiation 
at a redshift $z \approx 0.8$ \citep[see also][]{sato13}.
So we can speculate that the major part 
of the contaminating emission comes from approximately this distance or,
 more likely, from a range of intermediate distances corresponding to $z \in (0,1.5)$.

\section{Discussion}

The fact that the remote contaminating medium must be at a very low 
temperature (see Figure~\ref{fig:bb_observed_slopes}) 
can give clues as to the nature of the medium.
Obviously, hot intergalactic gas with its temperatures reaching $10^7$\,K 
can be disregarded as a candidate
for the contaminating agent, 
as well as dust in star-forming regions whose temperatures are 
too high, being of the order of 20\,K to 100\,K 
 \citep{galametz16}.  
This leaves us with the most likely contaminating ingredient being cold dust 
which for some time has been suspected to populate intergalactic space.  

\citet{eigenson38,eigenson49} and \citet{zwicky51, zwicky52, zwicky57}
were the first to notice the existence of intergalactic extinction due to dust
when studying the Coma Cluster of galaxies.
They have demonstrated 
that intergalactic dust could be detected by counting high-redshift objects
in the directions of lower-redshift clusters of galaxies. 
The presence of extragalactic dust 
was later traced by measuring the 
attenuation of distant background galaxies by
foreground galaxies \citep{gonzalez98, alton01}.     

The possibility of grey intergalactic extinction was debated just after
the discovery of the SN\,Ia dimming, see, e.g., the review by \citet{riess00} 
who wrote that 
{\sl ``dust which is greyer than Galactic-type dust could challenge the
cosmological interpretation of high-redshift SNe\,Ia''}. At the time of that
discussion, there was not sufficient evidence supporting such a possibility.
But more recently, the theories of interstellar and intergalactic dust 
have been substantially revised
\citep[see][]{voshchinnikov12, schultheis15, hutton15, vavrycuk19}.
It has become clear that at the peripheries of galaxies and possibly further away in the
intergalactic medium, the fraction of coarse-grain dust is larger than in the galactic disks,
which leads to this dust resembling the theoretical ``grey'' dust
that leaves little or no imprint on the spectral energy distribution of background sources.  
It also creates the long-known excess of radiation from some extragalactic objects in the 
far IR at $\lambda \approx 500\, \mu$m, which extends up to centimetre wavelengths and
which was confirmed and measured by the 
{\it Herschel} and {\it Planck} space observatories \citep{galliano11,ade11a}. 

In the 1990s, this excess was interpreted as an elevated spatial mass 
density of cold dust with temperatures of 4 to 7\,K \citep{reach95}. 
At that time such an interpretation was considered to be impossible.
However, the observed anticorrelation between 
the 500\,$\mu$m emission and the density of the medium \citep{galliano03,galliano05} 
supports this interpretation 
because, 
if the radius of dust grains 
in an environment with a constant spatial mass density
grows by an order of magnitude, then the number density of grains 
will be reduced by three orders of magnitude,
which would dramatically increase the transparency of the medium 
and would reduce the interactions of dust particles with radiation, rendering
them difficult to detect.
The angular sizes of the observed regions with the 500\,$\mu$m emission
range from 0.02$^\circ$ \citep{lisenfeld02} to 0.5$^\circ$ \citep{galliano11}.
So this emission would effectively distort the CMB power spectrum at the 
multipole moments $\ell \approx 360$ and higher, near the first
trough of the curves shown in Fig.\,\ref{fig:h},
where the effect of the CMB distortion on the calculated parameter $H_0$
is quite strong.

More observational evidence supporting our hypothesis comes from the 
directional dependence of the Universe acceleration parameter as estimated from 
the type Ia SNe data \citep{cai12, bernal17, colin19}, which is currently interpreted
as an artefact of us, observers, being located in a local bulk flow. However, it would be more 
logical to assume a non-homogeneous distribution of intergalactic dust rather than
anisotropy of Universe's acceleration.

We conclude that the mechanism proposed here for contamination of CMB radiation 
by some distant cold foreground emission can explain the 
discrepancy between the local measurements of $H_0$ and the {\it Planck}-derived 
value, without invoking assumptions that would require modifications of the standard cosmological
model or the standard model of particle physics.

\section*{Acknowledgements}
We would like to thank Prof. Mat Page, Dr. Alice Breeveld, 
Dr. Leslie Morrison for useful discussions
on the matters in this paper and an anonymous referee for detailed suggestions allowed to fix a few
mistakes in our calculations and to essentially improve the structure of our manuscript.

\end{document}